\documentclass[prl,aps,amsfonts,nofootinbib,notitlepage,longbibliography,10pt,superscriptaddress,twocolumn]{revtex4-1}

\usepackage{graphicx}% Include figure files
\usepackage{tabularx}
\usepackage{braket}
\usepackage{amssymb,mathtools}
\usepackage{amsmath}
\usepackage{bm}
\usepackage[driverfallback=dvipdfm]{hyperref}
\hypersetup{colorlinks=true,breaklinks,urlcolor=blue,linkcolor=blue,citecolor=blue}
\usepackage[normalem]{ulem}
\usepackage{xcolor}

\newcommand{\comment}[1]{}
\definecolor{ginger}{rgb}{0.69, 0.4, 0.0}

\usepackage{tikz}
\usetikzlibrary{arrows.meta,positioning,fit,calc}

\newcommand{\lr}[1]{\left( #1\right)}
\newcommand{\mlr}[1]{\left[ #1\right]}
\newcommand{\glr}[1]{\left\{ #1\right\}}
\newcommand{\alr}[1]{\left\langle #1\right\rangle}
\newcommand{\norm}[1]{\left\lVert#1\right\rVert}
\newcommand{\abs}[1]{\left\lvert#1\right\rvert}
\newcommand{\rarrow}{\quad \Rightarrow \quad}

\newcommand{\ee}{\mathrm{e}}

\newcommand{\tr}[1]{\mathrm{Tr}\lr{#1}}
\newcommand{\trace}{\mathrm{Tr}}

\newcommand{\order}{\mathrm{O}}

\newcommand{\cR}{\mathcal{R}}

\newcommand{\cZ}{\mathcal{Z}}

\newcommand{\cL}{\mathcal{L}}

\newcommand{\OO}{\mathcal{O}}

\newcommand{\bz}{{\bm{z}}}

\newcommand{\bs}{{\bm{s}}}
\newcommand{\pushright}[1]{\ifmeasuring@#1\else\omit\hfill$\displaystyle{#1}$\fi\ignorespaces}

\newcommand{\kett}[1]{\left|#1\right\rangle\mkern-9mu\left.\vphantom{#1}\right\rangle}

\makeatletter
\renewcommand{\p@subsection}{}
\renewcommand{\p@subsubsection}{}
\makeatother

\usepackage{amsthm}
\newtheorem{thm}{Theorem}
\begin{document} 

\title{Local redundancies are necessary for long-lived memories in finite spatial dimensions}
\author{Chao Yin}\email{chaoyin@stanford.edu}
\affiliation{Department of Physics, Stanford University, Stanford, California 94305, USA}

\author{Yaodong Li}\email{liyd@uw.edu}
\affiliation{Department of Mechanical Engineering \& Department of Physics, University of Washington, Seattle, WA 98195}

\author{Benedikt Placke}\email{benedikt.placke@physics.ox.ac.uk}
\affiliation{%
 Rudolf Peierls Centre for Theoretical Physics,
 University of Oxford,
 Clarendon Laboratory, Parks Road, Oxford, OX1 3PU, UK
}%

\author{Vedika Khemani}\email{vkhemani@stanford.edu}
\affiliation{Department of Physics, Stanford University, Stanford, California 94305, USA}

\date{\today}

\begin{abstract}
% We prove a no-go theorem for exponentially long-lived thermal memories, based on the structure of redundancies among Hamiltonian terms (parity-checks). 
% For any local stabilizer code in finite spatial dimensions without local redundancies, we prove that local heat-bath dynamics is \emph{fast mixing}, with $O(N)$ mixing time at any constant nonzero temperature, where $N$ is the number of qubits.
% For any local stabilizer code  in finite spatial dimensions without local redundancies, we prove that the mixing time of local thermalization dynamics is $O(N)$ at any constant nonzero temperature, where $N$ is the number of qubits.
% We establish this by proving a strong decay of correlations property of the thermal Gibbs state. Furthermore, we prove \emph{rapid mixing} with $\order(\log N)$ mixing time given a stronger condition on the induced redundancy patterns in suitably chosen regions, which applies to Haah's cubic code and generalizations, as well as three-dimensional layer codes. The absence of local redundancies precludes thermodynamic phase transitions in this setting, but does not necessarily rule out glassy dynamics. Indeed, glassiness can protect quantum information for exponentially long times in non-Euclidean quantum LDPC codes, even without an accompanying thermodynamic phase transition. Our results resolve the longstanding question,  exemplified by Haah's cubic code, of whether such a mechanism can also support passive memories in finite spatial dimensions. 

We prove a no-go theorem for exponentially long-lived thermal  memories, based on the structure of redundancies among  Hamiltonian terms. For local stabilizer Hamiltonians in finite spatial dimensions, the absence of local redundancies---relations in which the product of a small number of Hamiltonian terms equals the identity---is known to preclude bulk thermodynamic phase transitions. Yet glassy dynamics can still sustain  long-lived memories even without a thermodynamic  transition, as demonstrated by LDPC codes on non-Euclidean graphs. A longstanding question, exemplified by Haah's cubic code, is whether such a  mechanism can also support passive (quantum) memory in finite-dimensional Euclidean systems.
We show that this route fails for local stabilizer Hamiltonians without local redundancies: at any constant nonzero temperature, local heat-bath  dynamics is \emph{fast mixing}, with $O(N)$ mixing time, where $N$ is the number of qubits. We establish this by proving a \emph{strong decay of correlations} property of the thermal Gibbs state. We further prove \emph{rapid mixing} with  $O(\log N)$ mixing time under a stronger condition on induced redundancies in suitably chosen regions. This stronger result applies to several well-studied examples, including Haah's cubic code and its generalizations, as well as three-dimensional layer codes. Our work identifies redundancy structure as a key property for understanding thermal memories.

\end{abstract}

\maketitle

\noindent \emph{\bf Introduction.}---
Reliable storage of quantum information is a central challenge for present-day quantum computing. 
A particularly attractive solution is a passive quantum memory, in which information encoded in a low-temperature many-body state survives for exponentially long times without active intervention \cite{q_memory02,qmemory_rev16}. Stabilizer codes provide a natural setting, with their parity checks defining local Hamiltonian terms. A central question is what physical mechanisms can support passive memories, and what constraints limit their existence.

Slow relaxation dynamics---fundamentally necessary for memory---is often associated with thermodynamically ordered phases, suggesting that passive memory should accompany a finite-temperature bulk phase transition. For local stabilizer Hamiltonians in finite spatial dimensions,  such transitions require local redundancies among Hamiltonian terms, whose absence precludes singularities in the free-energy density~\cite{WeinsteinOrtizNussinov19,rakovszky2023physics,LDPC_memory_trivialF25}. Physically, these redundancies constrain the nature of excitations. In the one-dimensional Ising model, the absence of local redundancies produces point-like,   deconfined domain walls that proliferate at any nonzero temperature, precluding stable order. In contrast, in two dimensions, local redundancies constrain domain walls to form extended loops, whose free-energy cost grows with their size at sufficiently low temperatures, allowing stable order and a long-lived classical memory whose lifetime grows (stretched) exponentially with system size~\cite{Peierls36}. A similar contrast appears between the two-dimensional and four-dimensional toric codes, with the latter supporting a self-correcting quantum memory~\cite{q_memory02,qmemory_rev16,TC_fastmix09,AlickiHorodecki10}.

A long line of work, however, suggests that slow \emph{glassy dynamics} may offer an alternative route to exponentially long-lived memories, possibly without any thermodynamic transition. One motivation comes from kinetic constraints, which can strongly impede the motion of excitations even when they are point-like, leading to slow relaxation despite simple equilibrium thermodynamics. An early classical example is the Newman--Moore model, where such constraints lead to unusually slow relaxation~\cite{NewmanMoore99,NewmanMoore_Garrahan00}. Haah's cubic code, a celebrated candidate for self-correcting quantum memory in three dimensions, provides a quantum analogue of this idea~\cite{Haah_code11,Haahcode_Bravyi11}. More recently, layer codes have been constructed in three dimensions with polynomially growing energy barriers~\cite{layercode24, Gu_layer}. All three models lack local redundancies and display substantial slowing of dynamics over finite scales~\cite{Haah_code13,layercode_partial_mem_williamson_2026,Gu_layer}, but for a long time it remained unclear whether such slowing could persist asymptotically with growing system size, giving rise to genuinely glassy dynamics and  long-lived memory. In non-Euclidean settings, by contrast, this possibility is fully realized: certain classical and quantum LDPC codes on expander graphs have no local redundancies and an analytic free-energy density at all nonzero temperatures, yet are proven to exhibit a sharp dynamical glass transition and exponentially long-lived memory~\cite{MontanariSemerjian06,LDPC_memory_trivialF25,topoQSG24,PlackeSommersEtAl25}.

These developments leave open a central question: can glassy dynamics without local redundancies produce a self-correcting memory in finite-dimensional Euclidean systems? Here we show that it cannot. For local stabilizer Hamiltonians in finite spatial dimensions without local redundancies, we prove that local heat-bath  dynamics~\cite{Glauber63} is fast mixing at any fixed nonzero temperature, with mixing time $t_{\rm mix}=O(N)$ for $N$ qubits. Thus, the absence of local redundancies rules out not only passive memory, but also the existence of a glass phase  in this setting. We establish this by proving a \emph{strong decay of correlations} property~\cite{strong_decay_anyd24} in the thermal Gibbs state. Under a stronger condition on induced redundancies in local regions, we further prove rapid mixing, $t_{\rm mix}=O(\log N)$. This stronger result applies to all the aforementioned examples, including the Newman-Moore model, Haah's cubic code and its generalizations, as well as three-dimensional layer codes.

% These developments leave open whether glassy dynamics without redundancies can produce a self-correcting memory in finite-dimensional Euclidean space.
% In this work we show that it cannot.
% For local stabilizer Hamiltonians in $D$ dimensions with no local redundancies, evolving under Glauber dynamics \cite{Glauber63} at any nonzero temperature, we prove that the memory time is at most $O(N)$, see Theorem~\ref{thm:main}.
% Thus, in Euclidean space, local redundancies are necessary for self-correction, and glassiness alone is not enough.

Our result rules out passive memory in a large class of models in a unified manner and via a very simple to check common property. 
While several interesting candidates covered by our theorem have very recently been ruled out on a case-by-case basis~\cite{plaquet17,StengeleCahaCapelWarzel26,BartholdiHartarskyMitrofanov26},  here we uncover a unifying mechanism responsible for their failure.
We also significantly constrain the search space for passive memories compared to prior work \cite{BravyiTerhal08,qmemory_2dnogo13,3dqmemory_Yoshida11,fastmix_Ebarrier17,polynomial_Haah13}, with a particularly interesting case being codes which are local in three dimensions.
% self-correcting memories.
Within the chain complex formulation of CSS codes, checks and qubits form a three-term chain complex, and each layer of local redundancies extends it by one term.
Our result implies that a self-correcting quantum memory requires at least a five-term chain complex, to capture local redundancies in both $X$ and $Z$ checks, and embedding this structure in three dimensions is highly nontrivial, as realized recently in Ref.~\cite{3dqmemory_26}. Within this context, our results establish that such an embedding is fundamentally necessary.

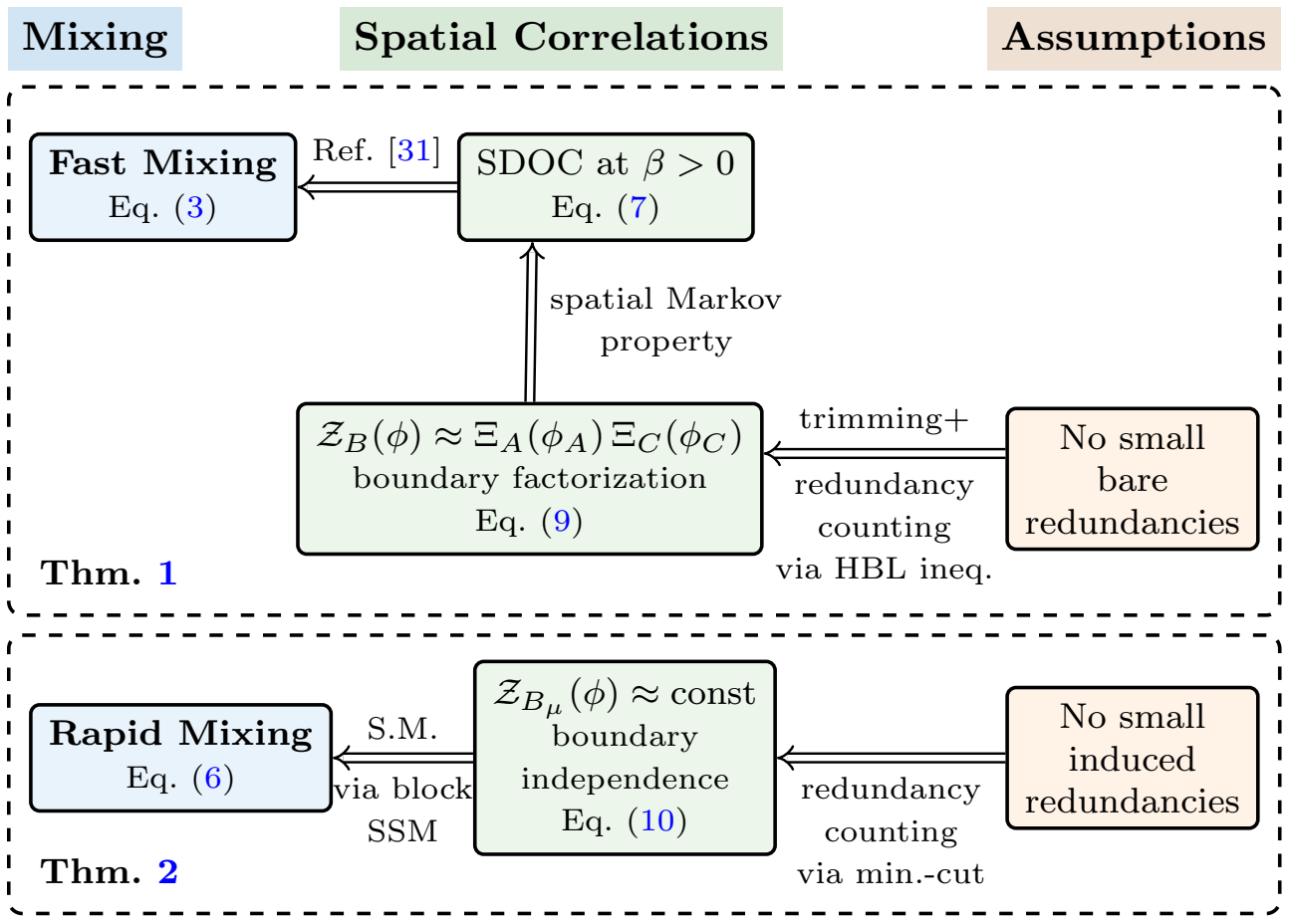
\begin{figure}
    \centering

\resizebox{\columnwidth}{!}{%
        % Color-only changes relative to the supplied TikZ.
% Node positions, styles, arrows, labels, and frames are untouched.
% TikZ automatically loads xcolor.

\begin{tikzpicture}[
    font=\footnotesize,
    box/.style={draw=black,thick,rectangle,rounded corners=2pt,align=center,inner xsep=4pt,inner ysep=3.5pt,minimum height=0.75cm},
    imp/.style={draw=black,double,double distance=1.4pt,-{Implies[length=4pt,width=5pt]},line width=0.55pt},
    coltitle/.style={font=\bfseries\small,align=center,anchor=south},
    thmlabel/.style={font=\bfseries\footnotesize},
    edgelabel/.style={font=\footnotesize,align=center,inner sep=1pt},
    dashframe/.style={draw=black,dashed,thick,rounded corners=3pt,inner sep=4.5pt}
]
\definecolor{flowTemporalHeader}{HTML}{D2E5F5}
\definecolor{flowSpatialHeader}{HTML}{D9E9D7}
\definecolor{flowAssumptionHeader}{HTML}{EEE1D3}
\definecolor{flowTemporal}{HTML}{E9F3FC}
\definecolor{flowSpatial}{HTML}{ECF6EA}
\definecolor{flowAssumption}{HTML}{FFF3E7}
\def\colOne{0}
\def\colTwo{3.00}
\def\colThree{7.50}
\node[box, fill=flowTemporal, minimum width=1.25cm] (fastmix)
    at (\colOne-4pt, 0)
    {\textbf{Fast Mixing}\\\scriptsize Eq.~\eqref{eq:fast_mix}};
\node[box, fill=flowSpatial, minimum width=2.05cm] (sdoc)
    at (\colTwo+0.35, 0)
    {SDOC at $\beta>0$\\\scriptsize Eq.~\eqref{eq:SDOC}};
\node[box, fill=flowSpatial, minimum width=2.5cm] (zb1)
    at (\colTwo-.25, -2.3)
    {$\cZ_B(\phi)\approx \Xi_A(\phi_A)\,\Xi_C(\phi_C)$\\
     \scriptsize boundary factorization\\
     \scriptsize Eq.~\eqref{eq:ZB-factorized}};
\node[box, fill=flowAssumption, minimum width=1.45cm] (bareR)
    at (\colThree, -2.3)
    {No small\\bare\\redundancies};
\draw[imp] (sdoc.west) -- (fastmix.east)
    node[midway, above=3pt, edgelabel] (refabove) {\scriptsize Ref.~\cite{SDOC26}};
\draw[imp] (zb1.north) -- ([xshift=-5.8mm]sdoc.south)
    node[midway, right=3pt, edgelabel]
        {\scriptsize spatial Markov \\\scriptsize property};
\draw[imp] ([yshift=2mm]bareR.west) -- ([yshift=2mm]zb1.east)
    node[midway, above=3pt, edgelabel] {\scriptsize trimming+}
    node[midway, below=3pt, edgelabel] (lieblabel)
        {\scriptsize redundancy\\\scriptsize counting\\\scriptsize via HBL ineq.};
\node[box, fill=flowTemporal, minimum width=1.25cm] (rapidmix)
    at (\colOne, -4.5)
    {\textbf{Rapid Mixing}\\\scriptsize Eq.~\eqref{eq:rapidmix}};
\node[box, fill=flowSpatial, minimum width=2.05cm] (zb2)
    at (\colTwo+0.5, -4.5)
    {$\cZ_{B_\mu}(\phi)\approx\mathrm{const}$\\
     \scriptsize boundary\\\scriptsize independence\\
     \scriptsize Eq.~\eqref{eq:ZZ-1<}};
\node[box, fill=flowAssumption, minimum width=1.45cm] (indR)
    at (\colThree, -4.5)
    {No small\\induced\\redundancies};
\draw[imp] (zb2.west) -- (rapidmix.east)
    node[midway, above=3pt, edgelabel] (smabove) {\scriptsize S.M.}
    node[midway, below=3pt, edgelabel] (smlabel) {\scriptsize via block\\\scriptsize SSM};
\draw[imp] (indR.west) -- (zb2.east)
    node[midway, below=3pt, edgelabel] (mincutlabel)
        {\scriptsize redundancy\\\scriptsize counting \\ \scriptsize via min.-cut};
% ---- frame section: unchanged geometry
\coordinate (shared-left) at (fastmix.west);
\coordinate (shared-right) at (bareR.east);
\coordinate (frame1-nw) at ($(shared-left |- sdoc.north) + (0, 0.20cm)$);
\coordinate (frame1-se) at ($(shared-right |- lieblabel.south) + (0, -1pt)$);
\coordinate (frame2-nw) at ($(shared-left |- zb2.north) + (0, 1pt)$);
\coordinate (frame2-se) at ($(shared-right |- mincutlabel.south) + (0, -1pt)$);
\node[dashframe, fit=(frame1-nw)(frame1-se)] (frame1) {};
\node[dashframe, fit=(frame2-nw)(frame2-se)] (frame2) {};
\node[thmlabel, anchor=south west] (thm1)
    at ($(frame1.south west) + (4pt, 4pt)$) {Thm.~\ref{thm:main}};
\node[thmlabel, anchor=south west] (thm2)
    at ($(frame2.south west) + (4pt, 4pt)$) {Thm.~\ref{thm:rapid}};
\node[coltitle, fill=flowTemporalHeader, anchor=south west]
    at ($(frame1.north west) + (0, 3pt)$) {Mixing};
\node[coltitle, fill=flowSpatialHeader]
    at ($(\colTwo, 0 |- frame1.north) + (0, 3pt)$) {Spatial Correlations};
\node[coltitle, fill=flowAssumptionHeader, anchor=south east]
    at ($(frame1.north east) + (0, 3pt)$) {Assumptions};
\end{tikzpicture}
    }
    \caption{Flow of implications for the main results of the paper. 
    Depending on the strength of assumptions on redundancies, we can conclude either fast or rapid temporal mixing, in both cases via showing an intermediate \emph{spatial} correlation decay property.}
    \label{fig:flowchart}
\end{figure}

% \begin{figure}
%     \centering
%     \includegraphics[width=\linewidth]{proof_flowchart.pdf}
%     \caption{Caption}
%     \label{fig:placeholder}
% \end{figure}

\noindent \emph{\bf Setting and  Results.}--- Consider a $d$-dimensional lattice with one qubit on each of its $N$ vertices.
Consider %\ben{a family of?} \chao{the results just apply to any one such Hamiltonian, so i prefer not to say a family; relatedly I have added a clarification of the big-O notation below.}\ben{but without defining a family, isn't eq (2) trivial? If I only consider one hamiltonian, then everything is $O(1)$, no?}\chao{No. See the statement of Theorem 1 in supp, which holds for every fixed $N$ as long as it is sufficiently large. I just want $t\ge c N$ for a constant $c$ determined by the other parameters of the problem.}\ben{I see, I guess what you are saying is that the constant are universal \emph{across Hamiltonians}, and only depend on stuff like the range of interaction.} \chao{yes.} 
stabilizer Hamiltonian \begin{equation}\label{eq:H=sum_s}
    H=-\sum_{i\in \mathcal{I}} (h_i - I)
\end{equation}
with Pauli strings $\{h_i:i\in \mathcal{I}\}$ commuting with each other and $I$ being the identity. The checks $h_i$ generate the stabilizer group $\mathcal{S}$. 
This includes classical codes where the $h_i$'s are strings of Pauli $Z$s, and quantum Calderbank-Shor-Steane (CSS) codes with each $h_i$ being a $Z$-string or an $X$-string. Each $h_i$ acts on a local region of size independent of $N$. We say $H$ has no local redundancies (i.e. relations) if for any nonempty set $R\subset \mathcal{I}$ of checks, \begin{equation}\label{eq:no_local_red}
    \prod_{i\in R}h_i = I \rarrow |R|\ge (\log N)^{d+1}
\end{equation}
i.e. grows at least as a large enough polylogarithm of $N$. %\vedika{what about log sized redundancies?} \vedika{specify locality of checks} \chao{ok changed to polylog, which is a slightly stronger bound on $R$ than supp. added one sentence above on locality.}\vedika{why large polynomial? $d+1$ isn't necessarily large?}

Let $\pi = \frac{\ee^{-\beta H}}{\tr{\ee^{-\beta H}} }$ be the thermal Gibbs state at inverse temperature $\beta$. Our main result is:
\begin{thm}[informal]\label{thm:main}
    For any stabilizer code in finite spatial dimensions with no local redundancies,
     single-site heat-bath
    Glauber dynamics 
    % \YL{do we need to specify details of Glauber, e.g. single-step versus heat-bath? I assume the latter}\chao{ok added, indeed people do not really use name ``Glauber dynamics'' for quantum settings, but i think at least ``heat-bath Glauber dynamics'' should be a valid name.} 
    at any constant temperature $\beta^{-1} > 0$ is fast mixing, with mixing time\footnote{In this paper, $f=\order(g)$ means $f\le c\cdot g$ for a constant $c$ determined by $d,\beta,\delta,\delta',c'$ and the Hamiltonian range.} 
    % \YL{``fast mixing''}
    \begin{equation}\label{eq:fast_mix}
        t_{\rm mix}=\order(N).
    \end{equation}
\end{thm}
% \YL{Emphasize that this holds regardless of boundary conditions? I guess boundary conditions can also be formulated in the form of Eq.(1).}\chao{yes as long as there are no local redundancies in $H$. emphasized below.}
% \vedika{is it true for any local detailed balance obeying dynamics? }\chao{it is a bit awkward to rigorously define ``any local DB dynamics'', because u need to rule out the dynamics of doing nothing. I.e. the definition would need a uniform local strength $>1$ condition. Although I expect the result is true given such an appropriate definition, most literature just state for particular dynamics anyway.}
Here the heat-bath Glauber dynamics \cite{Gibbssamp_commuting16} is a natural detailed-balanced local Lindbladian $\cL$ that, with unit rate, resamples each qubit to local thermal equilibrium determined by its neighbors, and the mixing time is defined by the smallest $t$ such that
\begin{equation}
    \norm{\ee^{t\cL}(\rho) - \pi}_1 \le \frac{1}{4}
\end{equation}
for any initial state $\rho$. In particular, Theorem \ref{thm:main} forbids exponential-time memories because all initial states approximately converge to $\pi$ after $t_{\rm mix}$. It applies to any boundary conditions including periodic ones, as long as \eqref{eq:no_local_red} holds. 

We also discuss a related result, which uses a stronger assumption on the structure of redundancies to conclude \emph{rapid} mixing [that is $t_{\rm mix}=\order(\log N)$]. The stronger assumption demands that small redundancies are absent even when restricting \eqref{eq:H=sum_s} to certain subsystems, and we informally call this the absence of small \emph{induced} redundancies (defined below).

\begin{thm}[informal]\label{thm:rapid}
    Suppose for any $L\ge 1$, there is a connected region $B_0$ with diameter $\le L$, such that for every lattice translation $B_\mu$ of $B_0$, every nontrivial induced redundancy $R\subset \mathcal{I}_B$ is large \begin{equation}\label{eq:R>condition}
        |R|>c'L^{\delta'}
    \end{equation}
    for some constants $c',\delta'>0$ independent of $L,B_\mu$. Then at any constant temperature, a constant-size heat-bath block dynamics has mixing time \begin{equation}\label{eq:rapidmix}
        t_{\rm mix}=\order(\log N).
    \end{equation}
\end{thm}
Here, the heat-bath block dynamics are a Lindbladian $\cL=\sum_\Lambda \cL_\Lambda$, where each $\cL_\Lambda$ locally resamples a region $\Lambda$ with constant diameter, rather than a single qubit in the Glauber dynamics before. More precisely, $\Lambda$ is chosen to be a union of some nearby $B_\mu$s at a given size $L$ depending on $\beta$.
% \ben{there is a relation between $B$ above and $\Lambda$ here that we should at least mention. I assume they are not the same?}\chao{yes, added one sentence.}
% Such jump operators are physical in nature\ben{sounds weird, perhaps say local, or physical?}\chao{ok}, 
The heat-bath dynamics is a natural generalization of the single-site one (with slightly increased non-locality),
and we expect the rapid mixing result could be generalized to the latter case.

% At the end of this letter, we also discuss the related Theorem \ref{thm:rapid}, which uses are stronger assumption on the structure of redundancies to conclude \emph{rapid} mixing [that is $t_{\rm mix}=\order(\log N)$]. The stronger assumption demands that small redundancies are absent even when restricting \eqref{eq:H=sum_s} to certain subsystems, and we informally call this the absence of small \emph{induced} redundancies.

Formal proofs of the two theorems are contained in the Supplemental Material (S.M.) \cite{SM}, and we sketch the proofs below.
Both proofs proceed by showing a form of spatial correlation decay as an intermediate step. We summarize the flow of implications for both theorems in Fig. \ref{fig:flowchart}, which may serve as a visual guide to the remainder of this letter.

% The intuition comes from correlation decay ...

% energy vs entropy ..

\begin{figure}
    \centering
    \includegraphics[width=0.98\linewidth]{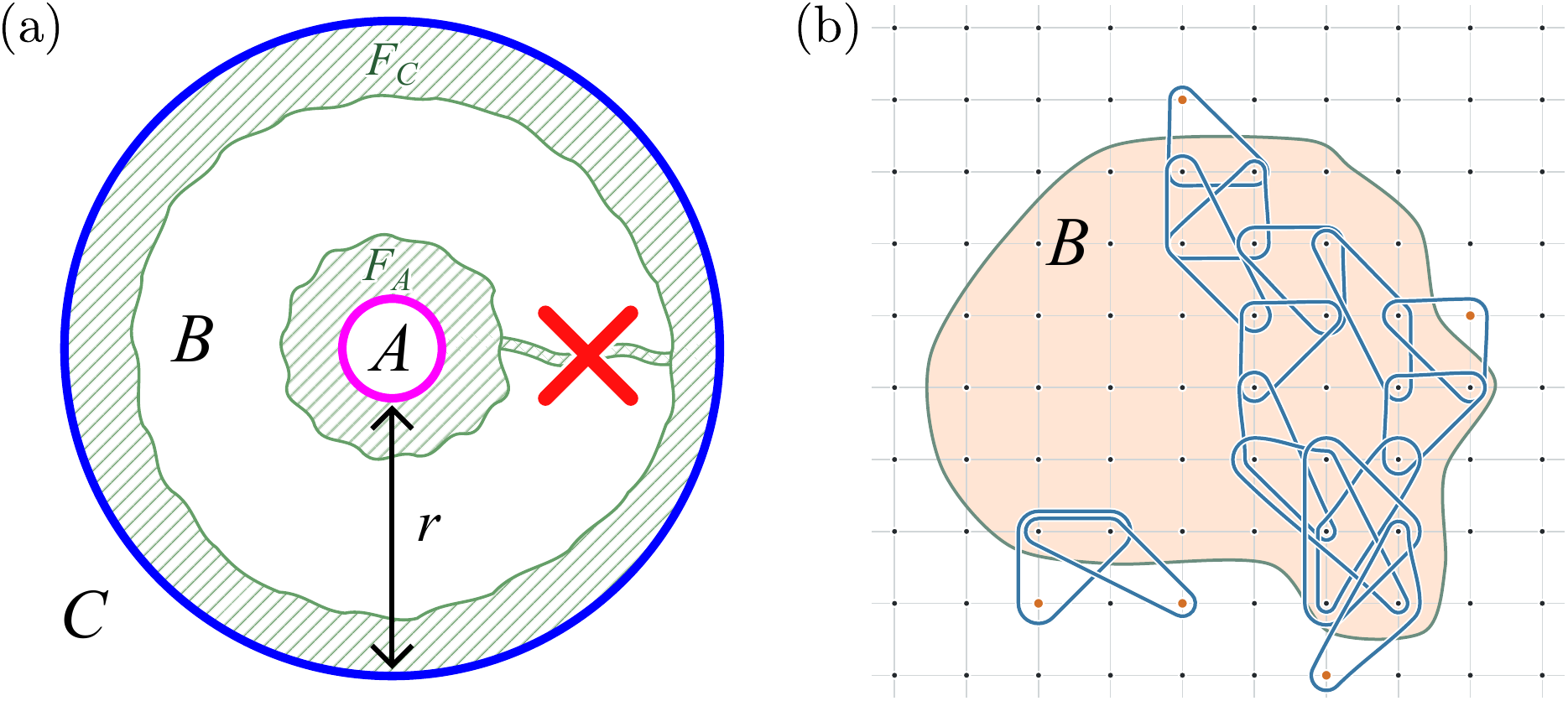}
    \caption{(a) For any stabilizer Hamiltonian without local redundancies at any constant temperature, we prove strong decay of correlations between regions $A,C$ with $B$ in between. This reduces to controlling a partition function \eqref{eq:ZB=fix} in $B$ with fixed states on $A,C$ and some fixed check values in $F=F_A\sqcup F_C$, which should be disconnected at the two boundaries. (b) Illustration of induced redundancies in any region $B$ that give rise to the partition function \eqref{eq:ZB=}, where each blue loop represents a check on the enclosed qubits (vertices). }
    \label{fig:ABC}
\end{figure}

\noindent \emph{\bf From spatial correlations to temporal mixing}---
Slow dynamics in physics is often associated with the notion of a nontrivial phase, and the latter in turn often implies some form of spatial long-range correlations. On a more technical level, this idea is phrased as the relation between temporal and spatial mixing \cite{Martinelli99,SSM_rev04,spectral_independence_rev23}: the relation between the time needed to reach the steady state and the form of spatial correlations in that steady state. 
The exact relation between the two is subtle, and in our two main results we use the absence of redundancies to show slightly different forms of spatial correlation decay, which will imply fast and rapid mixing respectively.

%
% Theorem~\ref{thm:main} follows from a short-range correlation property of $\pi$, named \textit{strong decay of correlations} (SDOC), which we now introduce.

\emph{Fast mixing from strong decay of correlations}---
The first condition we use, in Theorem \ref{thm:main}, is \emph{strong decay of correlations} (SDOC). We quantify correlation decay by considering a tripartition of the lattice into regions $A,B,C$ as shown in Fig.~\ref{fig:ABC}, with distance $r$ between $A,C$. 
In this case, we inspect the state on $A$ and, informally speaking, demand that it is independent from that on $C$. Formally we say that a state $\pi$ satisfies strong decay of correlations (SDOC) if there exists $c > 0$ such that 
\begin{equation}\label{eq:SDOC}
    \norm{\pi_{A|\phi_C} - \pi_A}_1 \lesssim \ee^{-cr},
\end{equation}
for any tripartition $ABC$ with $r\lesssim \log N$ \footnote{Decay up to this scale is sufficient for establishing fast mixing; Further, our proof method requires that there are no redundancies in $A\cup B$, which is why we require redundancies to grow polylogarithmically in \eqref{eq:no_local_red}.}, and any fixed configuration $\phi_C$ on $C$. Here, $\pi_A:=\trace_{A^{\rm c}}(\pi)$ with $A^{\rm c}$ the complement of $A$, and $\pi_{A|\phi_C}$ is the state on $A$ \emph{conditioned on} $C$.
For a classical probability distribution $\pi(\bz)$ of bitstrings $\bz$, the conditioned state is just the conditional distribution $\pi(\bz| \bz_C)$, where $\bz_C$ denotes the (fixed) configuration of bits in $C$. 
In the quantum case, the state on region $C$ is projected to $\phi_C$ and $\pi_{A| \phi_C}:=\trace_{A^{\rm c}}\lr{\frac{\bra{\phi_C}\pi \ket{\phi_C}}{\tr{\bra{\phi_C}\pi \ket{\phi_C}}}}$ is a postselected state on $A$. For stabilizer codes, we construct the conditioning states $\phi_C$ as follows. Let $\partial C$ denote the outer boundary of $C$, consisting of qubits outside $C$ that share a check with qubits inside $C$. Consider global stabilizers $O\in\mathcal S$ that act trivially on $\partial C$, but may have support elsewhere outside $C$. For each such $O$, let $O_C$ denote its restriction to $C$ such that $\OO=\OO_C\otimes \cdots$ where $\cdots$ do not act on $C$. We collect all those $O_C$ and define
$\mathcal S_C=\{O_C: O\in\mathcal S,\ O_{\partial C}=I\}$, which is a commuting Pauli group \cite{SM}.
% : since $O$ commutes with every check and acts trivially on $\partial C$, its restriction $O_C$ also commutes with every check

We choose $\phi_C$ from a basis of pure stabilizer states that are simultaneous eigenstates of $\mathcal S_C$ \footnote{We will also use $\partial A,\phi_A$, $\mathcal{S}_A$, $A^c$, and $\partial B,\phi_B$, $\mathcal{S}_B$, $B^c$ defined similarly.}.
Technically, to prove Theorem~\ref{thm:main} the decay in \eqref{eq:SDOC} may be relaxed to slightly weaker than exponential, and we may only require \eqref{eq:SDOC} to hold for $r$ that are sufficiently large compared to the size of $A$; 
% \ben{for what purposes, so far SDOC has not been used for anything} \chao{we can prove \eqref{eq:SDOC} only with this requirement, bc r has to be at least $\log|A|$ for the trimming procedure to work}
we leave these details to the S.M \cite{SM}.

% \vedika{Does SDOC need to hold for every pure stabilizer state that is a simultaneous eigenstate of $\mathcal S_C$, or only for states in a particular chosen basis? Since $\mathcal S_C$ need not be maximal, how is this basis specified?}

In companion work \cite{SDOC26} by some of the authors, it is shown that if SDOC holds at all temperatures, that is sufficient for fast mixing \eqref{eq:fast_mix} at any $\beta<\infty$. %We refer to \cite{SDOC26} for the proof and comparisons to previous approaches. 
The idea is that SDOC at all temperatures implies the Gibbs states at different temperatures are smoothly connected to each other.
% \ben{SDOC over some range implies that the states in this range are smoothly connected to each other within this range, presumably?} \chao{yes, if u mean the temperature case here is a special case of the general ``some range''}. 
More precisely, the canonical purification of the Gibbs state $\kett{\sqrt{\pi}}:=(\sqrt{\pi}\otimes I)\kett{\Omega}$, where $\kett{\Omega}$ is a maximally entangled state ($N$ Bell pairs) between two copies of the system, is connected to the infinite-temperature $\kett{I}$ by a quasi-local unitary (i.e. roughly speaking, a constant-depth local unitary circuit). In other words, $\kett{\sqrt{\pi}}$ is short-range entangled. On the other hand, the Glauber dynamics $\cL$ under a similarity transformation maps to a frustration-free local Hamiltonian $\mathcal{H}$ on the doubled space with $\kett{\sqrt{\pi}}$ being its ground state \cite{Gibbssamp_commuting16}. This short-range entanglement property of $\kett{\sqrt{\pi}}$ enables one to utilize 
% \YL{meaning of enhance unclear}\chao{ok changed.} 
local thermalization rates of $\cL$ to derive a constant global spectral gap of $\mathcal{H}$ and therefore $\mathcal{L}$, which implies fast mixing. In conclusion, to prove Theorem \ref{thm:main} it suffices to show SDOC at \emph{all} constant temperature $\beta$; note that this fails for models with local redundancies like the two-dimemsional Ising model.
% \YL{SDOC at any $\beta$, or all $\beta$?}\chao{they mean the same?}
% \YL{We were discussing for a while a correlation decay condition, and how that might lead to fast mixing via heat-bath dynamics. Is that irrelevant now because we're actually using SDOC?} \chao{i am not sure what correlation decay condition u are refering to? But anyway i think SDOC is conceptually the weakest one to imply fast mixing.}

\emph{From SDOC to buffer partition functions.}---
% \YL{how about ``A stronger condition than SDOC''?}\chao{ok accommodated}.}---
% \YL{It is somewhat confusing that SDOC is defined as a condition uniformly in all tripartitions $ABC$, but here we allow ourselves to choose nice tripartitions, or even trimming.
% I suppose the actual trimming procedure can be carried out successfully for all ABC, but it is not clear given this detour to ``nice'' ABCs.} \chao{yes trimming applies to all ABC. }
Rather than establishing SDOC directly, we use the spatial Markov property \cite{thermal_markov12} of Gibbs states to reduce the problem to the boundary dependence of a partition function on the intervening buffer region $B$.
% While establishing SDOC directly can be tricky,
% we can exploit the spatial Markov property of Gibbs states to reduce the problem to the boundary dependence of a partition function on the intervening buffer region $B$.
% conclude \eqref{eq:SDOC} by inspecting the boundary dependence of the buffer region $B$. 
Informally, the Markov property means that any correlations between $A$ and $C$ have to be mediated through $B$. This motivates studying the partition function on $B$, conditioned on both $A$ and $C$
\begin{equation}\label{eq:ZB=}
\cZ_B(\phi):=\tr{\bra{\phi_A}\bra{\phi_C}\ee^{-\beta H_B}\ket{\phi_A}\ket{\phi_C}}
\end{equation} 
where $H_B=-\sum_{i\in \mathcal{I}_B} (h_i - I)$ with $\mathcal{I}_B\subset \mathcal{I}$ being checks touching $B$, and we will label $\phi_A$ and $\phi_C$ jointly by $\phi$. 
To show Theorem \ref{thm:main}, we will use  the absence of redundancies to show that the boundary dependence of $\cZ_B$ approximately factorizes, i.e. there exists a constant $c>0$ such that
\begin{equation}\label{eq:ZB-factorized}
    \cZ_B(\phi)= \Xi_A(\phi_A)\cdot \Xi_C(\phi_C)\cdot\mlr{1+\order(\ee^{-cr})}
\end{equation}
for all $\phi_{A, C}$ 
% \ben{is the constant universal over tripartitions?}\chao{sure, added ``constant''. Since \eqref{eq:SDOC} already has $c$ and is said to hold for all tripartition there, maybe we do not need to stress that again.}.
A short explicit calculation \cite{SM} using the Markov property then shows that this factorization \eqref{eq:ZB-factorized} implies SDOC \eqref{eq:SDOC} and hence fast mixing~\eqref{eq:fast_mix}. Establishing that \eqref{eq:ZB-factorized} follows from the absence of local redundancies is the core of proving our main result, Theorem \ref{thm:main}.

{
In some cases---including important examples such as Haah's
cubic code and layer codes---we can
establish a stronger form of boundary insensitivity,
albeit for a special family of regions rather than
the annular buffer $B$ considered above.
For the regions $B_\mu$ appearing in
Theorem~\ref{thm:rapid}, we show that the partition
function is approximately \emph{independent} of boundary
conditions,
\begin{equation}
\mathcal Z_{B_\mu}(\phi)
=\overline{\mathcal Z}_{B_\mu}
\left[1+\order\left(e^{-L^{\alpha}}\right)\right],
\label{eq:ZZ-1<}
\end{equation}
where $\overline{\mathcal Z}_{B_\mu}$ is independent of the stabilizer state
$\phi$ outside $B_\mu$ (similar to $\ket{\phi_A} \ket{\phi_C}$ in \eqref{eq:ZB=}),
$\alpha>0$ is a constant, and $L$ characterizes the size of these regions.
This condition is closely related to strong spatial
mixing (SSM) in the classical literature, which controls
the influence of boundary perturbations on distant
regions. A celebrated result is that SSM implies rapid
mixing of Glauber dynamics~\cite{Martinelli99,SSM_rev04}.
A related condition has also been formulated for CSS
codes~\cite{logSobolev_CSS26}. The regions $B_\mu$ can be assembled into screening regions \cite{plaquet17},
allowing us to establish SSM \cite{SM} 
% \ben{one some sufficiently large block $\Lambda$?}\chao{We did not define SSM explicitly, so ``for some block'' does not make much sense.}. 
Adapting the classical argument then gives rapid mixing for appropriate heat-bath
block updates. Establishing that \eqref{eq:ZZ-1<} follows from a stronger assumption on local redundancies (i.e. the lack of small induced redundancies) is therefore the core of proving our second main result, Theorem \ref{thm:rapid}.
}

In the S.M. we explicitly discuss an example of a system that fulfills the factorization condition \eqref{eq:ZB-factorized} without fulfilling the stronger boundary-independence condition \eqref{eq:ZZ-1<} for certain choices of $B$. The system is a modified 2d Ising model, where each nearest-neighbor bond $\alr{u,v}$ is ``gauged'' by an extra qubit on that bond: $H_{\rm gIsing}=-\sum_{\alr{u,v}}Z_u Z_v Z_{\alr{u,v}}$. This model is redundancy-free because each term has its own gauge qubit. However one can induce order along the boundary by carefully choosing boundary conditions, a phenomenon known as \emph{boundary order} \cite{Martinelli99,SDOC26}.

We will now show that the boundary dependence of $\cZ_B$ is controlled by a class of check relations that we call \emph{induced redundancies}. To see this, we expand the Boltzmann weights of the commuting checks in ~\eqref{eq:ZB=}, using $e^{\beta(h_i-I)}=e^{-\beta}\cosh\beta(1+h_i\tanh\beta)$. Since we trace over $B$, only terms in which the product of checks acts trivially on $B$ survive. This motivates the notion of induced redundancies: subsets $R\subset \mathcal{I}_B$ of checks for which the product $\bm{h}_R:=\prod_{i\in R}h_i$ has no support in $B$. {
Unlike a bare redundancy, $\bm{h}_R$ need not be the identity globally; it can have support in $B^c$, specifically on the boundary $\partial B$ (see Fig.~\ref{fig:ABC}(b) for an illustration). For the annulus geometry,  $\bm{h}_R\in\mathcal S_A\otimes\mathcal S_C$. } 
The boundary states $\phi$ are therefore eigenstates of $\bm{h}_R$, with eigenvalue $x_R=\pm1$. Performing the trace over $B$ then gives
\begin{equation}\label{eq:ZB=sumR}
    \frac{\cZ_B(\phi)}{\mathcal{N}} =  1+\sum_{R\in \cR_0:R\neq \emptyset}x_R (\tanh \beta)^{|R|}
\end{equation}
where $\mathcal N$ is a normalization, $\mathcal R_0$ is the set of all induced redundancies, and $|R|$ is the number of checks in $R$. Thus, the boundary dependence of $\cZ_B$ is controlled by the sizes and number of induced redundancies: each contribution is suppressed by $(\tanh\beta)^{|R|}$, but the combined effect depends on how many such redundancies exist of each size. {In particular, \emph{small} induced redundancies can contribute terms that remain significant even as the size of $B$ grows, with signs $x_R$ that depend on the boundary conditions. Such contributions can obstruct the stronger boundary-independence condition~\eqref{eq:ZZ-1<}, even if they are localized near one boundary and therefore potentially compatible with the weaker factorization~\eqref{eq:ZB-factorized}.}

We first sketch the proof of Theorem~2, where
the absence of small induced redundancies on
suitable regions permits a direct counting argument to control their contribution to $\cZ_B$.
We then return to the more general Theorem~1,
where small induced redundancies cannot be excluded
and a trimming procedure is needed.

{\noindent \emph{\bf Proof sketch of Theorem 2.}---
Equation~\eqref{eq:ZB=sumR} suggests a direct route to proving Theorem~\ref{thm:rapid}: the assumption~\eqref{eq:R>condition} ensures that induced redundancies are large in the regions $B_\mu$, suppressing their individual contributions to $\cZ_{B_\mu}$. However, we must also control their \emph{entropy}, i.e., how many redundancies exist of a given size. 
In the End Matter, we bound this entropy using a minimal-cut
argument: each redundancy is uniquely determined by its
smallest cross-sections along suitably chosen spatial slices,
so we can count these cuts instead of the redundancies themselves.
Counting these cuts bounds the total contribution of induced redundancies, establishing \eqref{eq:ZZ-1<}. As described above, this gives SSM on suitable blocks and hence rapid mixing~\cite{SM}. We also verify the required assumption for Haah's code in the End Matter, and for its generalizations and three-dimensional layer codes in the S.M.~\cite{SM}.}

% \ben{I'd be inclined to just state theorem 2 here and get over with it, and then have the proof of theorem 1 be the rest of the main text.}\chao{sure}
% Before stating this in more detail, we now explain how to get around assuming the absence of small induced redundancies to show Theorem \ref{thm:main}. Our strategy for this is, roughly speaking to ``smoothen'' the boundary condition by a trimming procedure.

% Indeed, we will show later that the existence of such good boundaries, which we identify in interesting models, implies not just fast but rapid mixing of thermalization dynamics \YL{defined with respect to these boundaries.}
% For general models \YL{for which we wish to establish SDOC for an arbitrary tripartition $A|B|C$}, however, it is unclear how to get rid of small induced redundancies in generality.
% % for all $A|B|C$.

% % \YL{Might need to adjust the previous few sentences accordingly, to be consistent with our narrative at the beginning of this subsection.}

\noindent \emph{\bf Proof sketch for Theorem 1.}---
We now sketch the proof of our main result Theorem \ref{thm:main}. 
% with full details given in the S.M. \cite{SM}. 
As discussed above, it is sufficient to show that the absence of local redundancies  \eqref{eq:no_local_red} implies a factorization of the buffer partition function as in \eqref{eq:ZB-factorized}. {Small induced redundancies in \eqref{eq:ZB=sumR}
need not, by themselves, obstruct factorization:
if all induced redundancies were built from checks
localized near either boundary, their contributions to
$\cZ_B$ would factorize between $A$ and $C$.
The difficulty arises when there are also large redundancies
spanning the buffer: combining these with different small
redundancies can lead to a proliferation of many distinct
spanning redundancies, potentially overwhelming the
factor $(\tanh\beta)^{|R|}$ suppressing each term
in \eqref{eq:ZB=sumR}.}
% To separate these effects, we fix the values $s$ of a
% carefully chosen set $F$ of checks and decompose
% $\cZ_B$ into conditioned partition functions,
To deal with the effect of small induced redundancies in \eqref{eq:ZB=sumR}, we rewrite this equation as a sum over partition functions with a specific subset of checks $F$ fixed to a bitstring $\bs$. That is we write $\cZ_B(\phi) = \sum_{\bs}\ee^{-\beta(2|\bs|-|F|)} \cZ_B(\phi,\bs)$ where $|\bs|$ is the Hamming weight, and
\begin{equation}\label{eq:ZB=fix}
\cZ_B(\phi,\bs):=\tr{\bra{\phi_A}\bra{\phi_C}Q_{\bs}\ee^{-\beta H_{B,F}}\ket{\phi_A}\ket{\phi_C}}
\end{equation}
where $Q_{\bs}=\prod_{i\in F}\frac{I+(1-2s_i)h_i}{2}$ is the projector onto the syndrome $\bs$, and $H_{B,F}$ only contains checks $\mathcal{I}_B\setminus F$. The object in \eqref{eq:ZB=fix} can be viewed as the partition function of a 'trimmed' model $H_{B,F}$ where the checks in $F$ have been removed from the Hamiltonian and their values fixed as part of the boundary conditions. Note that in general, not all values of $\phi$ and $\bs$ are consistent, and $\cZ = 0$ for inconsistent pairs. %\ben{zero or one, or some other constant?} \ben{perhaps more details on this?}\chao{zero. just because $Q_{\bs}\ket{\phi_A \phi_C}=0$}.

Below, we show that we can choose this set $F$ carefully,
such that it disconnects into two separate pieces $F_A$
and $F_C$ attached to $A$ and $C$ (see Fig.~\ref{fig:ABC}),
and there exists a constant $\overline{\cZ}'$ such that
\begin{align}
\left|\frac{\cZ_B(\phi,\bs)}{\overline{\cZ}'}-1\right|
\le \sum_{R\in\cR:R\neq\emptyset}
(\tanh\beta)^{|R|}
\lesssim \ee^{-cr},
\label{eq:ZBs<}
\end{align}
for all consistent pairs $\phi,\bs$.
Here $\cR=\{R\setminus F:R\in\cR_0\}$ denotes the
\emph{remaining} induced redundancies.
The first inequality is a direct calculation, while the second bound follows by controlling both the sizes
and number of remaining redundancies via a counting argument, as discussed below. Given this, the only dependence on $\phi$ in
$\cZ_B\propto\sum_{\bs}\ee^{-2\beta|\bs|}
\cZ_B(\phi,\bs)$ comes through the consistency
condition on $\phi$ and $\bs$.
Because the allowed syndromes in $F_A,F_C$ are independent,
this implies the factorization in \eqref{eq:ZB-factorized}.

\emph{Trimming redundancies}---
What's left to show is how to choose the set $F$. We do this via an iterative procedure. Intuitively, we add `small' induced redundancies to $F$ until $\mathcal{I}_B\setminus F$ only contains large redundancies. Then, a similar expansion as in \eqref{eq:ZB=sumR} but for $\cZ_B(\phi, \bs)$ can be used to show \eqref{eq:ZBs<}.
One challenge when choosing $F$ iteratively is to show that this procedure terminates before $F$ connects the boundaries of $A$ and $B$.
This is because removing checks in $F$ can reduce the support of an initially large induced redundancy $R$, leaving $R\setminus F$ small. Therefore, we perform a greedy trimming procedure as follows. We first present a simplified version that illustrates the idea, and elaborate on the full procedure after.
We first pick among $\cR_0$ the smallest induced redundancy $R_1\neq \emptyset$ and put it into $F$. We then search for the next smallest induced redundancy in the updated $\cR_1=\{R\setminus R_1:R\in \cR_0\}$ and attach it to $F$, and so on. We set a size cutoff $K$ so that the procedure terminates at some step $n$ when all induced redundancies are large $|R|\ge K$ for $\emptyset\neq R\in \cR$, where $\cR:=\cR_n$.

As mentioned above, we need the constructed $F$ to disconnected and   attached to each boundary separately, i.e. the procedure should terminate before $F$ grows to connect $A$ and $C$. Here we use the linear space structure of $\cR_0$: if $R,R'\in \cR_0$ then $R+ R'$ where $+$ means the symmetric difference is also an induced redundancy. Different $R$s have different operators $\bm{h}_R$ in $\partial B$, because otherwise the sum of different $R$s becomes a bare redundancy, which by assumption, does not exist in $B$. As a result, the dimension $\mathrm{dim}(\cR_0)=\order(|\partial B|)$. 
Furthermore, the updated $\cR_k$s continue to be linear spaces. The size of $F$ grows by at most $K$ at each step of the procedure; since the linear space decreases by one at each step, the total number of steps $n\le \mathrm{dim}(\cR_0)$ so that \begin{equation}\label{eq:F<}
    |F|\le nK=\order(K|\partial B|),
\end{equation}
i.e. it cannot take up the full region $B$ for $K\ll r$.
Eq. \eqref{eq:F<} alone, however, does not exclude the possibility that $F$ connects $A,C$ via a thin region. However, a refined argument in the S.M ~\cite{SM} also rules out this possibility, because a thin tunnel like Fig.~\ref{fig:ABC} would require much larger linear-space dimension than its base at the boundary can support. More precisely, one can show that $F$ cannot penetrate deep in the bulk: \begin{equation}\label{eq:F_close}
    \mathrm{dist}(i,B^{\rm c})\lesssim K,\qquad \forall i\in F.
\end{equation} 
We choose $K\sim r$ for $F$ to factorize into $F_A,F_C$.

The trimming procedure above is only for illustrative purposes. In the actual procedure used in the proof~\cite{SM}, we trim away a \emph{linear subspace} of induced redundancies at each step instead of just one, with $K\sim r$ now being the cutoff on the \emph{average} size. This improves the minimal-$R$ bound $|R|\ge K$ above to \begin{equation}\label{eq:S>Kdim}
    |S|\ge K\cdot \mathrm{dim}(\cR_S),\quad \forall S\subset \mathcal{I}_B\setminus F
\end{equation}
where $\cR_S$ is the subspace of all $R\in \cR$ supported in $S$, while \eqref{eq:F_close} still holds. 
% \YL{perhaps be more specific with $K \sim r$ versus $K \sim r / \log r$.}\chao{i'd prefer not to because otherwise every $\ee^{-cr}$ should be $\ee^{-cr/\log r}$ which looks a bit tedious.}
%
% \emph{Bounding the number of induced redundancies.}---
Eq. \eqref{eq:S>Kdim} turns out to be a strong condition because it controls not just the smallest $R$ but also the ``entropy'' of the induced redundancies by the dimension counting. A direct application of the vector space H\"{o}lder–Brascamp–Lieb inequality
\cite{HBLineq_Lieb76,HBLineq_Tao08,HBLineq_Tao10,HBL_ineq} then yields \eqref{eq:ZBs<}. SDOC \eqref{eq:SDOC} then follows, which concludes the proof of Theorem \ref{thm:main}.

\emph{Outlook.}--- We have shown that local redundancies are necessary for exponential-time thermal memories based on stabilizer codes in finite spatial dimensions. It remains an open question whether our rapid mixing results for specific models in this class extend more generally. At the same time, not all local redundancies support thermal memory, motivating the study of thermal states through their redundancy structure. Intriguingly, recent work \cite{redun_free_quantum26} introduces redundancy-free models with \emph{non-commuting} Pauli Hamiltonians that may realize nontrivial phases at low temperatures. Together, these observations suggest that the interplay between non-commutativity and redundancy is central to understanding thermal states of general quantum many-body systems.

\section{Acknowledgements}
We would like to thank Jeongwan Haah, Nicholas O'Dea, Rahul Sahay, David Long and Tibor Rakovszky for useful discussions. C.Y. and Y.L were partially supported by the US Department of Energy, Office of Science under Award No. DE-SC0019380 and by the Stanford Q-FARM Bloch Postdoctoral Fellowship in Quantum Science and Engineering. VK acknowledges support from the Packard Fellowship in Science and Engineering and the Office of Naval Research Young Investigator Program (ONR YIP) under Award Number N00014-24-1-2098.

\emph{AI Disclosure:} The authors conceived this project and formulated all its key conceptual ideas and hypotheses. ChatGPT was used to assist in developing the proofs and in writing the manuscript. The authors take full responsibility for the correctness of the paper.

\bibliography{biblio}

\section{End Matter}

\begin{figure}
    \centering
    \includegraphics[width=0.6\linewidth]{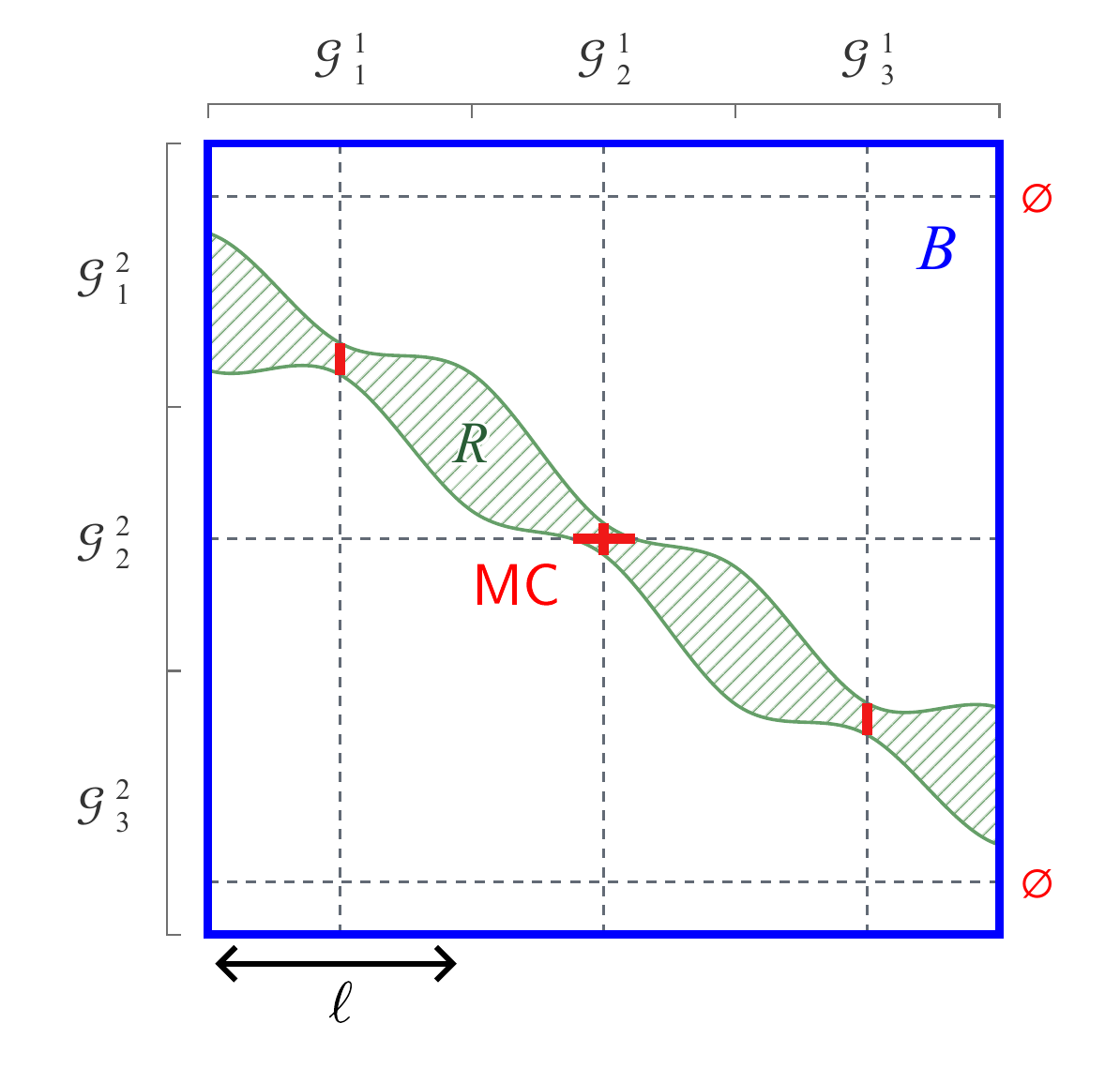}
    \caption{Illustration of the minimal cut for an induced redundancy $R$. We choose $\ell$ so that the minimal cut uniquely determines $R$.}
    \label{fig:MC}
\end{figure}

\emph{Counting large induced redundancies.}---
The main challenge in proving Theorem 2 is that to control \eqref{eq:ZB=sumR}, we need a counting argument that a typical $R\in \cR_0$ is large even if the worst-case bound \eqref{eq:R>condition} has a very small $\delta'$.  
We count $R$ by their support on a set of ``cuts'', defined as follows. Let $\alpha=1,\cdots,d$ be the directions in the Cartesian coordinate, the checks in region $B$ can be separated into ``hyperplanes'' with normal direction $\alpha$ \begin{equation}
    \mathcal{I}_B = \Pi^\alpha_1 \sqcup \Pi^\alpha_2 \sqcup\cdots \sqcup \Pi^\alpha_{L'}
\end{equation}
($L'\le L$)
such that only checks in adjacent planes can overlap. We put the planes into disjoint groups: $\mathcal{G}^\alpha_1= \{\Pi^\alpha_1,\cdots, \Pi^\alpha_{\ell}\}$ is the first group, $\mathcal{G}^\alpha_2= \{\Pi^\alpha_{\ell+1},\cdots, \Pi^\alpha_{2\ell}\}$ is the second group, and so on. The last group contains $\ge \ell$ elements. There are $g_{\rm max}\le L$ groups for each direction. Each group spans distance $\Theta(\ell)$ in direction $\alpha$. For each $R\in \cR_0$, it is cut by the planes with intersections $S^\alpha_k=\Pi^\alpha_k\cap R$. We define the \emph{minimal cut} $\{\mathsf{MC}^\alpha(R): \alpha=1,\cdots,d\}$ of $R$ by finding the smallest $S^\alpha_k$ in each group: \begin{equation}
    \mathsf{MC}^\alpha(R) = \cup_{k:S^\alpha_k \text{ is smallest in its group}} S^\alpha_k.
\end{equation}
We pick an arbitrary one if there are multiple choices in the group. See Fig.~\ref{fig:MC} as an illustration. In the connectivity graph of the checks, any connected component of $R\in \cR_0$ is also an induced redundancy, so \eqref{eq:R>condition} also bounds each connected component of $R$. By choosing \begin{equation}\label{eq:l=L^delta/d}
    \ell = c_\ell L^{\delta'/d}
\end{equation}
for a sufficiently small constant $c_\ell>0$, there must be one group $\mathcal{G}^\alpha_g$ of planes that all intersect with $R$, such that $|\mathsf{MC}^\alpha(R)|\ge 1 $ for that direction $\alpha$. This implies \begin{equation}
    \label{eq:MC>l}\sum_{\alpha=1}^d|\mathsf{MC}^\alpha|\ge 1
\end{equation}
for every valid minimal cut.

On the other hand, given the minimal cut $\{\mathsf{MC}^\alpha\}$, there is at most one $R\in \cR_0$ such that $\mathsf{MC}^\alpha(R)=\mathsf{MC}^\alpha$: Suppose there are two $R,R'$ with the same minimal cut, then $R+R'\in \cR_0$ has no support on the minimal cut so decomposes to connected components each contained in a small box of linear size $\sim \ell$. With a sufficiently small constant $c_\ell$ in \eqref{eq:l=L^delta/d}, each box cannot contain any nontrivial $R$ from \eqref{eq:R>condition}, which implies $R+R'=\emptyset\Rightarrow R=R'$.

Since the minimal cut determine a unique $R=R(\{\mathsf{MC}^\alpha\})$, we sum over the minimal cuts instead \begin{align}
    &\abs{\frac{\cZ_B}{\overline{\cZ}}-1} \le \sum_{\{\mathsf{MC}^\alpha\} } \lr{\tanh\beta}^{|R(\{\mathsf{MC}^\alpha\})|} \nonumber\\
    &\le \sum_{\{\mathsf{MC}^\alpha\} } \lr{\tanh\beta}^{\ell \frac{1}{d}\sum_{\alpha=1}^d|\mathsf{MC}^\alpha|} \nonumber\\
    &\le  \sum_{\{m^\alpha_g\}:\,\sum_{\alpha,g}m^\alpha_g\ge 1 } \sum_{\{k^\alpha_g\} } \sum_{S^\alpha_{k_g}\subset \Pi^\alpha_{k_g},|S^\alpha_{k_g}|= m^\alpha_g} \epsilon^{\sum_{\alpha,g}m^\alpha_g}  \nonumber\\
    &\le -1+ \sum_{\{m^\alpha_g\ge 0\} } \epsilon^{\sum_{\alpha,g}m^\alpha_g}  \prod_{\alpha,g}\ell^{\mathbb{I}\mlr{m^\alpha_g>0}} {c_d L^{d-1} \choose m^\alpha_g} \nonumber\\
&= -1+ \prod_{\alpha,g} \mlr{1+\ell \sum_{m=1}^{c_d L^{d-1}} {c_d L^{d-1} \choose m} \epsilon^m } \nonumber\\
&=-1+\glr{1+\ell\mlr{(1+\epsilon)^{c_d L^{d-1}} -1}}^{d g_{\rm max}}. \label{eq:cut_long<}
\end{align}
Here the second line uses $\abs{R(\{\mathsf{MC}^\alpha\})}\ge \ell |\mathsf{MC}^\alpha|$ that follows from the definition of the minimal cut being minimal in each group of planes. The third line first sums over the sizes $m^\alpha_k=|S^\alpha_k|\ge 0$ of the minimal cuts per plane, and then choices of the planes $\{k^\alpha_g\}$, and finally $S^\alpha_k$ with the given sizes on those planes. All $\alpha=1,\cdots,d$ and $g=1,\cdots,g_{\rm max}$ implicitly, and we used \eqref{eq:MC>l} and $\epsilon:=\lr{\tanh\beta}^{\frac{\ell}{d}}$. The fourth line bounds the last two sums, where $m^\alpha_g=0$ ($>0$) corresponds to one ($\ell$) choice(s) of the plane in the group shown by the indicator function $\mathbb{I}\mlr{m^\alpha_g>0}\in\{0,1\}$. It also introduces a constant $c_d$ such that
$|\Pi^\alpha_k|\le c_d L^{d-1}$,
and uses $-1$ to cancel the $m^\alpha_g\equiv 0$ contribution. %The second-to-last line of \eqref{eq} is because different $\alpha,g$ is independent.
Since $\epsilon$ is exponentially small in $L$ from \eqref{eq:l=L^delta/d}, 
\eqref{eq:cut_long<} becomes \begin{equation}\label{eq:ZZ-1<exp}
    \abs{\frac{\cZ_B}{\overline{\cZ}}-1}  \le \lr{1+\ell \epsilon\cdot 2c_d L^{d-1}}^{d g_{\rm max}}-1\le\ee^{-L^{\frac{\delta'}{2d}}}
\end{equation}
at sufficiently large constant $L$, where we have used $g_{\rm max}\le L$ and absorbed polynomial factors into $\epsilon$. As shown in SM, \eqref{eq:ZZ-1<exp} implies rapid mixing and proves Theorem \ref{thm:rapid}.

\begin{figure}
    \centering
    \includegraphics[width=0.8\linewidth]{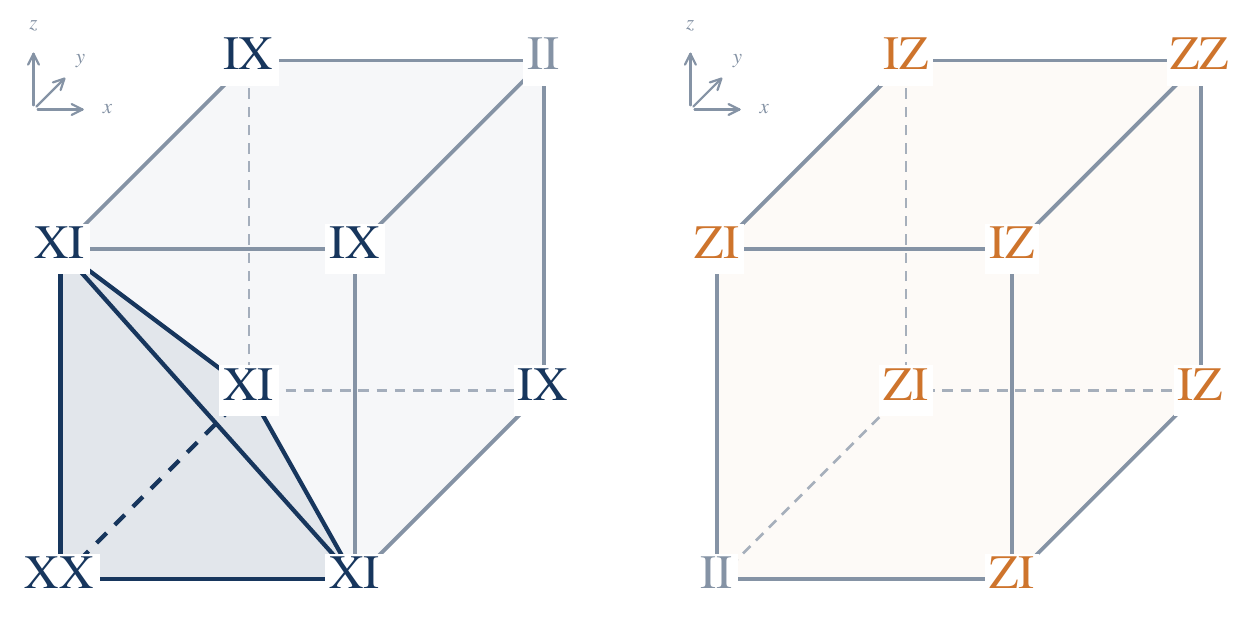}
    \caption{Checks of Haah's code. The shaded tetrahedron acts on type-$1$ qubits.}
    \label{fig:Haah}
\end{figure}

\begin{figure}
    \centering
    \includegraphics[width=1.0\linewidth]{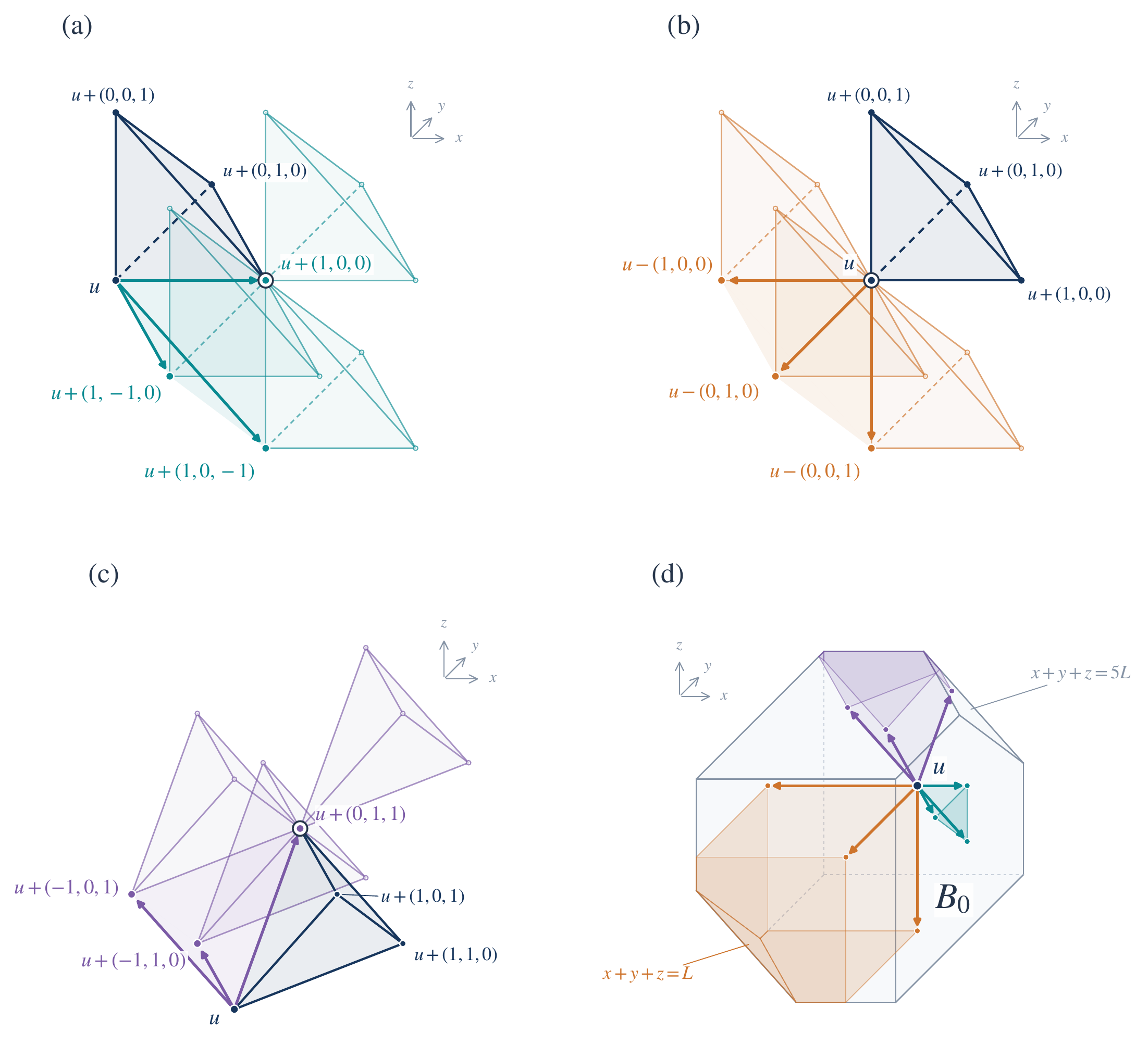}
    \caption{For Haah's code, induced redundancies grow in directions constrained by the arrows, which enforces them to scale with the region $B_0$ in (d).}
    \label{fig:haah_red}
\end{figure}

\emph{Haah's cubic code.}---
%Haah's cubic code \cite{Haah_code11,Haah_code13} was proposed as a candidate for quantum memory due to its fractonic dynamics and large energy barriers. Although  It 
The model is defined on the 3d cubic lattice where each vertex contains two qubits. The code is a CSS code $H=H^X + H^Z$ where $H^{X}$ ($H^Z$) contains only Pauli-$X$ ($Z$) strings as checks, and we have deleted an overall identity in $H$ comparing to \eqref{eq:H=sum_s}. The code is translation invariant with one $X$-check and one $Z$-check acting on each cube of the lattice shown in Fig.~\ref{fig:Haah}. We say the check is anchored at $u=(x_u,y_u,z_u)$ if it acts on the cube with $u$ being the vertex of it with smallest $x_u+y_u+z_u$ (i.e. the bottom-left corner in Fig.~\ref{fig:Haah}). 
We aim to show that \eqref{eq:R>condition} holds with constants $\delta'=1,c'$ for the region \begin{equation}\label{eq:B0_Haah}
    B_0 = \{u\in V: x_u,y_u,z_u\in[0,2L], L\le |x_u+y_u+z_u|\le 5L\}
\end{equation}  
shown in Fig.~\ref{fig:haah_red}(c), which is a cube with two corners cut out. By translation invariance and a constant rescaling of $L$, this proves rapid mixing for Haah's code from Theorem \ref{thm:rapid}; in SM this also generalizes to prime qudits, including the random code construction proposed in \cite{rand_local_code26}. We remark that the absence of topological order in the cubic code thermal state has been proven via dealing with redundancies \cite{Haahcode_trivial_Yoshida17}.

We focus on redundancies formed by $X$-checks alone because $Z$-checks are independent and related to $X$-checks by an inversion symmetry. Since each site contains two qubits that we call type $1$ and $2$, an induced redundancy $R=R^{(1)}\sqcup R^{(2)}$ has to have vanishing support for each $R^{(1)}\cap B_0=R^{(2)}\cap B_0=0$ separately. We first consider the relations among the tetrahedral part in Fig.~\ref{fig:Haah} for $R^{(1)}$. The idea for \eqref{eq:R>condition} is that an induced redundancy has to ``grow'' in certain directions, as shown in Fig.~\ref{fig:haah_red}. If a check (grey in Fig.~\ref{fig:haah_red}(a)) supported in $B_0$ anchored at $u$ is included in an induced redundancy, there must be another check included to cancel its support on $u+(1,0,0)$: There are three such choices shown in green in Fig.~\ref{fig:haah_red}(a), where we have arrows pointing from $u$ to those anchors forming a cone shape. For any choice of the three green checks, if it is supported in $B_0$ then there should be a check further to the right canceling its right-most vertex, and so on. As a result, we draw three green rays coming out of $u$ in Fig.~\ref{fig:haah_red}(d), and there must be a connected set of checks contained in the green cone that are included in the induced redundancy. 

The above arguments for canceling the $+(1,0,0)$ support also hold for the other three vertices in the tetrahedron, where two of them are related by symmetries and the last one of them for canceling the anchor is shown in Fig.~\ref{fig:haah_red}(b). For type-$2$ qubits with a different shaped tetrahedron in the check, there are also four cones emitting out of the vertex $u$, where Fig.~\ref{fig:haah_red}(c) shows one example. As a result, for any nontrivial redundancy $R\subset \mathcal{I}_{B_0}$, \begin{equation}\label{eq:R>haah}
    |R|\ge \min_{u\in B_0} \sum_{\alpha\in\{0,1,\cdots,7\} } \mathrm{dist}_\alpha(u,\partial B_0)
\end{equation}
where $\mathrm{dist}_1(u,\partial B_0)$ for example is the smallest number of vertices in the green region in Fig.~\ref{fig:haah_red}(c) that connects $u$ to the boundary $\partial B_0$, and $\alpha\in \{0,1,2,3\}$ corresponds to the four cones for type-$1$ qubits. For shape \eqref{eq:B0_Haah}, one can verify that for any $u\in B_0$ except for a small region around $(L,0,0)$, at least one of the four type-$1$ cones has to stretch in length $\sim L$ for all its three rays, leading to $\mathrm{dist}_\alpha(u,\partial B_0)\sim L$. On the other hand, if $u$ is close to $(L,0,0)$, the purple cone in Fig.~\ref{fig:haah_red}(c) stretches in length $\sim L$ because that cone never crosses the $x+y+z=L$ cut. \eqref{eq:R>haah} therefore establishes \eqref{eq:R>condition} for $\delta'=1$.

% attach supplement
\clearpage
\begingroup

% supp.pdf is 8.5 x 11 in
\pdfpagewidth=612bp
\pdfpageheight=792bp

% IMPORTANT: remove pdfTeX's normal 1-inch offset
\pdfhorigin=0pt
\pdfvorigin=0pt

\newcount\supppage
\supppage=1

\loop
  \pdfximage page \supppage {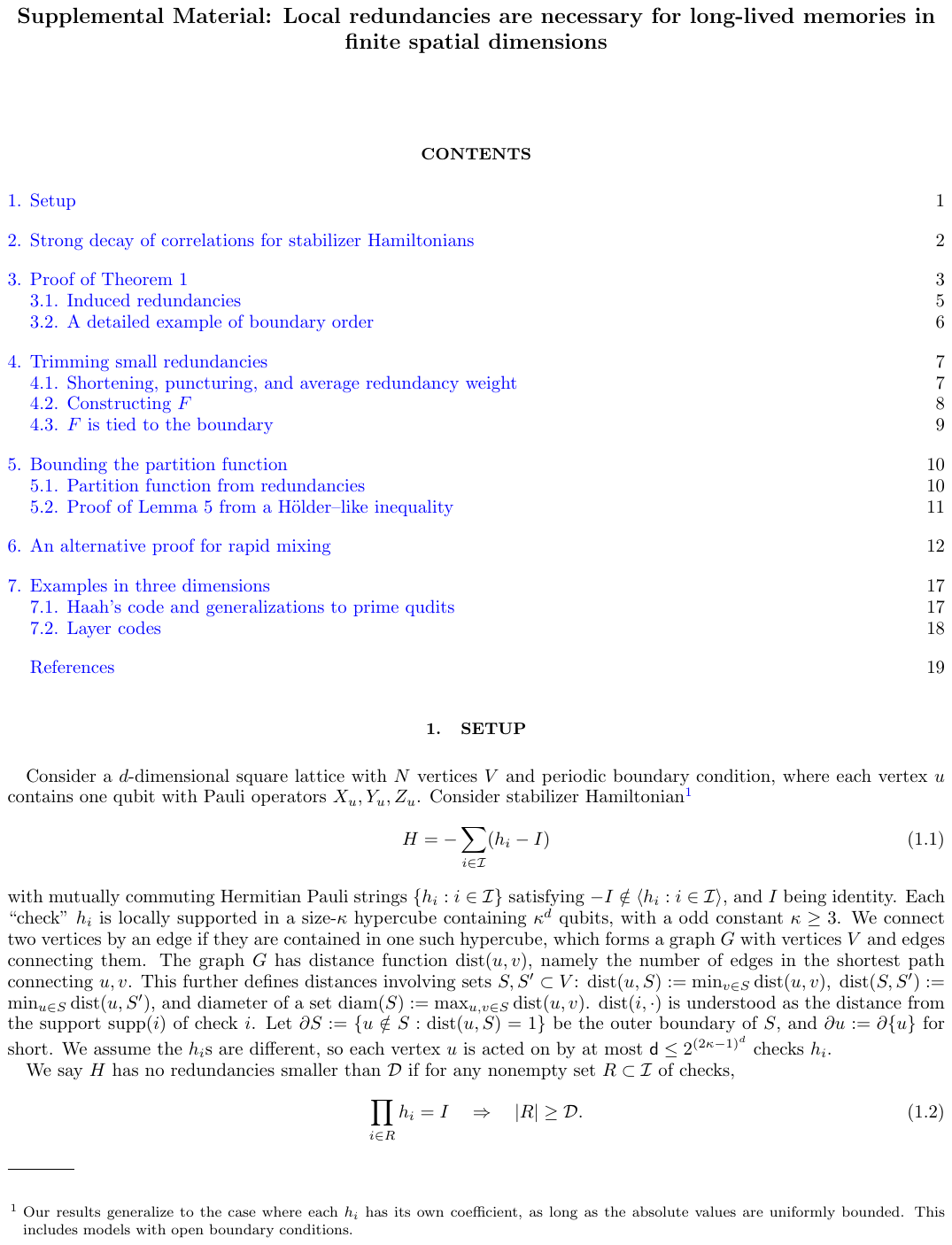}%
  \shipout\vbox to 792bp{%
    \offinterlineskip
    \hbox to 612bp{%
      \pdfrefximage\pdflastximage
      \hss
    }%
    \vss
  }%
  \advance\supppage by 1
\ifnum\supppage<21
\repeat

\endgroup

\end{document}